\documentclass[aps,prd,preprint,superscriptaddress,tightenlines,nofootinbib]{revtex4}

\usepackage{graphicx}
\usepackage{dcolumn}
\usepackage{bm}
\def\Journal#1&#2&#3(#4){#1{\bf #2}, #3 (#4)}

\def\NPB{Nucl.  Phys.  {\bf B}}
\def\PLB{Phys.  Lett.  {\bf B}}
\def\PRL{Phys.  Rev.  Lett.  }
\def\PRD{Phys.  Rev.  {\bf D}}

\def\etal{{\it et al.}}

\begin{document}
\preprint{CLEO CONF 04-06}
\preprint{ICHEP04 ABS10-0753}

\title{\bf Exclusive multi-body Hadronic Decays of the $\psi (2S)$}
\thanks{Submitted to the International Conference on High Energy Physics, 
August 2004, Beijing}

\author{Z.~Li}
\author{A.~Lopez}
\author{H.~Mendez}
\author{J.~Ramirez}
\affiliation{University of Puerto Rico, Mayaguez, Puerto Rico 00681}
\author{G.~S.~Huang}
\author{D.~H.~Miller}
\author{V.~Pavlunin}
\author{B.~Sanghi}
\author{E.~I.~Shibata}
\author{I.~P.~J.~Shipsey}
\affiliation{Purdue University, West Lafayette, Indiana 47907}
% ...
\collaboration{CLEO Collaboration} 
\noaffiliation

\date{\today}

\begin{abstract} 
Using data accumulated with the CLEO detector 
corresponding to an integrated luminosity of $\cal{L}$=5.46~pb$^{-1}$ on the 
peak of the $\psi(2S)$ and 20.46~pb$^{-1}$ at $\sqrt{s}$=3.67~GeV, we report 
preliminary branching fraction measurements for seven new decay modes 
of the $\psi(2S)$ 
($\eta 3\pi$, $\eta^\prime 3\pi$, $2(K^+ K^-)$, $p \bar p K^+ K^-$, 
$\Lambda \bar \Lambda \pi^+ \pi^-$, $\Lambda \bar p K^+$, and
$\Lambda \bar p  K^+ \pi^+ \pi^-$) and
more precise measurements of nine previously measured modes 
($2(\pi^+ \pi^-)$, $2(\pi^+ \pi^-) \pi^0$, $\omega \pi^+ \pi^-$, 
$K^+ K^- \pi^+ \pi^-$, $\phi \pi^+ \pi^-$, $\omega K^+ K^-$, $\phi K^+ K^-$, 
$p \bar p \pi^+ \pi^-$, and $\Lambda \bar \Lambda$). 
We also include a study of
$\omega p \bar p$ and obtain an improved upper limit 
for $\phi p \bar p$. Results are compared, where possible, with the 
corresponding $J/\psi$ branching ratios to test the 12\% rule.

\end{abstract}

\maketitle

In perturbative QCD the states $J/\psi$ and $\psi(2S)$ are 
non-relativistic bound
states of a charm and an anti-charm quark. The decays of these states are expected to be
dominated by the annihilation of the constituent $c\bar{c}$ into three
gluons. The partial width for the decays into an exclusive hadronic state, $h$,
is expected to be proportional to the square of the 
$c\bar{c}$ wave function overlap at the origin, which is well
determined from the leptonic width~\cite{PDG}. Since the strong coupling 
constant, $\alpha_s$, is not very
different at the $J/\psi$ and $\psi(2S)$ masses, it is expected that 
for any state $h$ the
 $J/\psi$ and $\psi(2S)$ branching ratios are related by~\cite{RULE}

\begin{equation}
Q_h=\frac{{\cal B}(\psi(2S)\to h)}{{\cal B}(J/\psi\to h)}
\approx
\frac{{\cal B}(\psi(2S)\to \ell^+\ell^-)}{{\cal B}(J/\psi\to\ell^+\ell^-)}= (12.7 \pm 0.5)\%,
\end{equation}
where ${\cal B}$ denotes a branching fraction, $h$ is a particular hadronic final
state, and the leptonic branching fractions are taken from the PDG~\cite{PDG}.
This relation  is sometimes called
\lq \lq the 12\% rule''. Modest deviations from the rule are expected~\cite{GULI}. 
Although the rule works well for some specific decay modes of the 
$\psi(2S)$, it fails spectacularly for $\psi(2S)$ decays to final states 
consisting of one vector and one pseudoscalar meson, such as $\rho\pi$. 

Values of $Q_h$ have been measured for a wide variety of final states~\cite{PDG,MOREBES,HARRIS}.
Most recently CLEO has measured additional $1^-0^-$ final states at the $\psi(2S)$ and 
the continuum~\cite{BHHMK}, including the first observation of  $\psi(2S) \rightarrow \rho\pi$.
A recent review~\cite{GULI} of relevant theory and experiment 
concludes that current theoretical explanations are unsatisfactory.
Clearly more experimental results are desirable. This paper presents 
measurements of the following new decay modes of the $\psi(2S)$: 
$\eta 3\pi$, $\eta^\prime 3\pi$, $2(K^+ K^-)$, $p \bar p K^+ K^-$,
$\Lambda \bar \Lambda \pi^+ \pi^-$, $\Lambda \bar p K^+$,
$\Lambda \bar p  K^+ \pi^+ \pi^-$,
and more precise measurements of these previously
measured modes: 
$2(\pi^+ \pi^-)$, $2(\pi^+ \pi^-) \pi^0$, $\omega \pi^+ \pi^-$,
$K^+ K^- \pi^+ \pi^-$, $\phi \pi^+ \pi^-$, $\omega K^+ K^-$, $\phi K^+ K^-$,
$p \bar p \pi^+ \pi^-$, and $\Lambda \bar \Lambda$.
We also measure $ \omega p \bar p $
and obtain an improved upper limit for $\phi p \bar p$.
Where applicable, the inclusion of charge conjugate modes is implied.
Eleven of the modes we study have been previously observed at the $J/\psi$.

The data sample used in this analysis is obtained at the $\psi(2S)$ and the nearby continuum
in $e^+e^-$ collisions produced by the Cornell Electron Storage Ring (CESR) and acquired with
the CLEO detector. 
The CLEO~III detector~\cite{cleoiiidetector} 
features a solid angle
coverage for charged and neutral particles of 93\%.
The charged particle tracking system, operating in a
1.0~T magnetic field along the beam axis, achieves 
a momentum resolution of $\sim$0.6\% at
$p=1$~GeV/$c$. The calorimeter attains a photon 
energy resolution of 2.2\% at $E_\gamma=1$~GeV and 5\% at 100~MeV.
Two particle identification systems, one based on energy loss ($dE/dx$) in 
the drift chamber and the other a ring-imaging Cherenkov (RICH) 
detector, are used together to separate $K^\pm$ from $\pi^\pm$. 
The combined $dE/dx$-RICH particle identification procedure has
efficiencies exceeding 90\% and misidentification rates below 5\%
for both $\pi^\pm$ and $K^\pm$.

Half of the $\psi(2S)$ data and all
the continuum data were taken after
a transition to the CLEO-c~\cite{YELLOWBOOK} detector configuration, in which
CLEO~III's silicon-strip vertex detector was replaced with a six-layer
all-stereo drift chamber. 
The two detector configurations also correspond
to different accelerator lattices: the former with
a single wiggler magnet and a center-of-mass
energy spread of 1.5~MeV, the latter 
(CESR-c~\cite{YELLOWBOOK}) with
six wiggler magnets and an energy spread of 2.3~MeV. 

The integrated luminosity ($\cal{L}$) of the datasets was measured 
using $\gamma\gamma$
events in the calorimeter~\cite{LUMINS}. Event counts were normalized with 
a Monte Carlo (MC) simulation based on the Babayaga~\cite{BBY} event 
generator combined with GEANT-based~\cite{GEANT} detector modeling. 
The datasets consist of  $\cal{L}$=5.46~pb$^{-1}$ on the 
peak of the $\psi(2S)$ (2.57~pb$^{-1}$ for CLEO~III, 
2.89~pb$^{-1}$ for CLEO-c) and 20.46~pb$^{-1}$ at $\sqrt{s}$=3.67~GeV
(all CLEO-c). The nominal scale factor used to normalize continuum yields 
to $\psi(2S)$ data is $f_{\rm nom}=0.2645\pm0.004$, 
and is determined from the integrated luminosities of the data sets corrected for the $1/s$ 
dependence of the cross section, where the error is from the relative luminosity uncertainty. 
The actual $f$ used for each mode also corrects 
for the small differences in efficiency between the $\psi(2S)$ 
and continuum data samples.

Standard requirements are used to select charged particles 
reconstructed in the tracking system and photon candidates in the 
CsI calorimeter. We require tracks of charged particles  
to have momenta $p>100$~MeV 
and to satisfy $|\cos\theta|<0.90$, where $\theta$ is the polar angle 
with respect to the $e^+$ direction.
Each photon candidate satisfies  $E_\gamma>30$~MeV and is more than 8\,cm
away from the projections of tracks into the calorimeter. 
Particle identification from $dE/dx$ and the RICH is used on all charged
particle candidates. 
Pions, kaons, and protons
must be positively and uniquely identified. That is: pions must not satisfy 
interpretation as kaons or protons, and kaons and protons obey similar requirements.
Charged particles must not be identified as electrons using criteria based 
on momentum, calorimeter energy deposition, and $dE/dx$.

The invariant mass of the decay products from the following particles 
must lie within limits determined from MC studies:  
$\pi^0~(M_{\gamma \gamma}=120-150 {\rm ~MeV})$, 
$\eta ~(M_{\gamma \gamma}=500-580 {\rm  ~MeV})$,
$\eta ~(M_{\pi^+ \pi^- \pi^0}=530-565 {\rm ~MeV})$, 
$\omega ~(M_{\pi^+ \pi^- \pi^0}=740-820 {\rm ~MeV}$
[$M_{\pi^+ \pi^- \pi^0}=760-800 {\rm ~MeV}$ 
for the $\omega p \bar p$ final state]), 
$\phi ~(M_{K^+ K^-} = 1.00-1.04 {\rm ~GeV})$, 
and 
$\Lambda ~(M_{p \pi^-} = 1.1136-1.1180 {\rm ~GeV})$.
For $\pi^0 \rightarrow \gamma \gamma$ and $\eta \rightarrow \gamma \gamma$ candidates
in events with more than two photons, the combination giving a mass closest to the
known  $\pi^0$ or $\eta$
mass is chosen, and a  kinematically constrained fit to the known parent mass is used.
Fake $\pi^0{\rm 's}$ and $\eta{\rm 's}$ are suppressed with lateral electromagnetic
shower profile restrictions.  
For $\eta \rightarrow \pi^+ \pi^- \pi^0$ and $\omega \rightarrow \pi^+ \pi^- \pi^0$,
the $\pi^0$ is selected as described above, and then combined with all possible combinations 
of two oppositely charged pions choosing the combination that is closest to the $\eta (\omega)$ mass.
A kinematically constrained fit is used for neither of these modes, 
nor for $\phi{\rm's}$ or for $\Lambda{\rm's}$.  
For $\Lambda \rightarrow p \pi^-$, a fit of the $p \pi^-$ trajectories to a common vertex 
separated from the $e^+ e^-$ interaction ellipsoid is made. 
Contamination from $K_S$ decays is eliminated primarily by the energy and momentum 
requirements imposed on the event, and by particle identification.   

Energy and momentum conservation requirements are imposed
on the reconstructed final state hadrons, 
which have momentum $p_i$ and combined 
measured energy $E_{\rm vis}$. We require the measured scaled energy 
$E_{\rm vis}$/$E_{\rm cm}$ 
be consistent with unity within experimental resolution, which varies by final state. 
We require $|\Sigma {\bf p_i}|/E_{\rm cm}< 0.02$. Together these requirements
suppress backgrounds with missing energy or incorrect mass assignments. 
The experimental resolutions are
smaller than 1\% in scaled energy and 2\% in scaled momentum.

For the final states $2(\pi^+\pi^-)\pi^0$,
$\omega\pi^+\pi^-$, and $\omega K^+K^-$, an additional cut is applied to 
remove a background of radiative events. When the photon is
combined with a low-energy photon signal, it can imitate a $\pi^0$.  
We require $(E_{\rm 4 tracks}+E_\gamma)$/$E_{\rm cm} < 0.995$, 
where $E_\gamma$ is the energy of the highest energy photon.

In order to compute $Q_h$ in modes with two or more charged pions,
two $\pi^0{\rm's}$ or an $\eta$, it
is necessary to remove  the contribution 
from the transitions $\psi(2S) \rightarrow J/ \psi X$, where $X= \pi^+ \pi^-$, $\pi^0 \pi^0$,
or $\eta$~\cite{note}. Accordingly we reject events in  
which the mass of any of the following falls within the range 
$3.05<m<3.15$~GeV: 
the two highest momentum oppositely charged tracks, the recoil mass against the two lowest
momentum oppositely charged tracks, or the mass recoiling against 
the 2$\pi^0{\rm's}$ or $\eta$.

 
For every  final state, a signal selection range in  
$E_{\rm vis}$/$E_{\rm cm}$ is determined by Monte Carlo simulation,
and a sideband selection range is defined to measure background.
Final states with the intermediate $\eta$, $\omega$, or  $ \phi$ particles 
must satisfy a scaled energy signal selection range requirement identical to the corresponding mode
without the intermediate particle.
For example, the scaled energy signal selection range is the same
for  $\phi K^+K^-$ and $K^+K^-K^+K^-$.
For final states with an  $\eta$, $\omega$, or $ \phi$, the event yield 
is determined
from signal and sideband selection ranges of the intermediate particle mass. 
The scaled energy and  mass signal and sideband selection ranges for
$\psi(2S)$ and continuum data are identical, and 
are listed by mode in Tables \ref{tab:xtot} and \ref{tab:resm}.

\begin{table}[!h]
\caption{Signal and sideband selection ranges for the scaled
energy, $E_{\rm vis}/E_{\rm cm}$, by mode.}
\begin{center}
\begin{tabular}{|c|c|c|}  \hline
   mode                        & signal & sideband  \\ \hline
$2(\pi^+ \pi^-)$               & 0.98-1.02 & 0.96-0.98, 1.02-1.04 \\
$2(\pi^+ \pi^-)\pi^0$          & 0.98-1.02 & 0.96-0.98, 1.02-1.04 \\
$K^+K^-\pi^+\pi^-$             & 0.99-1.01 & 0.98-0.99, 1.01-1.02 \\
$2(K^+K^-)$                    & 0.99-1.01 & 0.98-0.99, 1.01-1.02 \\
$p \bar{p} \pi^+ \pi^-$        & 0.99-1.01 & 0.98-0.99, 1.01-1.02 \\
$p \bar{p} K^+ K^-$            & 0.99-1.01 & 0.98-0.99, 1.01-1.02 \\
$\Lambda \bar\Lambda$          & 0.99-1.01 & 0.98-0.99, 1.01-1.02 \\
$\Lambda\bar\Lambda\pi^+\pi^-$ & 0.99-1.01 & 0.98-0.99, 1.01-1.02 \\
$\Lambda \bar{p} K^+$          & 0.99-1.01 & 0.98-0.99, 1.01-1.02 \\
$\Lambda\bar{p}K^+\pi^+\pi^-$  & 0.99-1.01 & 0.98-0.99, 1.01-1.02 \\ \hline
\end{tabular}
\label{tab:xtot}
\end{center}
\end{table}

\begin{table}[!h]
\caption{ The mass signal and sideband selection ranges for modes with 
an $\eta$, $\omega$, or $\phi$. Modes with these particles must satisfy 
a scaled energy signal selection requirement identical to the corresponding 
mode without the intermediate particle. 
}
\begin{center}
\begin{tabular}{|c|c|c|c|}  \hline
   mode                       &    Variable      & signal (GeV) & sideband (GeV) \\ \hline
$\omega \pi^+ \pi^-$          & $m_{3\pi}$       & 0.74-0.82 & 0.70-0.74, 0.82-0.86 \\
$\omega K^+ K^-$              & $m_{3\pi}$       & 0.74-0.82 & 0.70-0.74, 0.82-0.86 \\
$\omega p \bar{p}$            & $m_{3\pi}$       & 0.76-0.80 & 0.74-0.76, 0.80-0.82 \\ \hline
$\eta^{\to\gamma\gamma} 3\pi$              
                              &$m_{\gamma\gamma}$& 0.50-0.58 & 0.455-0.50,0.58-0.615 \\
$\eta^{\to\pi^+\pi^-\pi^0} 3\pi$              
                              & $m_{3\pi}$       &0.530-0.565& 0.51-0.53,0.565-0.58 \\
$\eta^\prime 3\pi$      &$m_{\pi\pi\gamma\gamma}$&0.945-0.970& 0.9325-0.945,0.97-0.9825 \\ \hline
$\phi \pi^+ \pi^-$            & $m_{K^+K^-}$     & 1.00-1.04 & 1.04-1.08 \\
$\phi K^+ K^-$                & $m_{K^+K^-}$     & 1.00-1.04 & 1.04-1.08 \\
$\phi p \bar{p}$              & $m_{K^+K^-}$     & 1.00-1.04 & 1.04-1.08 \\ \hline
\end{tabular}
\label{tab:resm}
\end{center}
\end{table}
\clearpage

\begin{figure}[htbp]
  \centering
  \includegraphics[height=0.9\textheight,bb=33 92 521 717,clip=true]
   {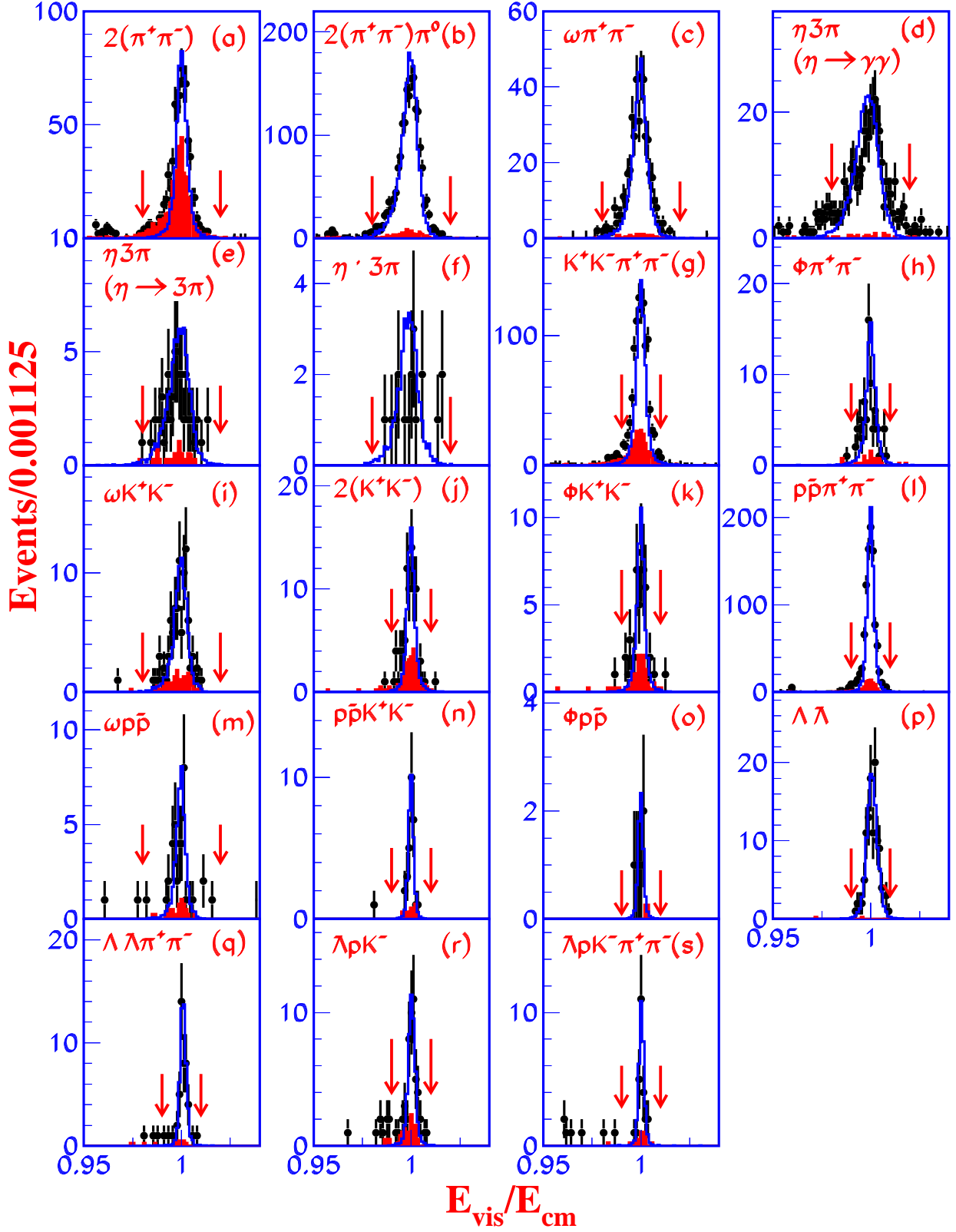}
  \caption{ Distributions of the scaled visible energy $E_{\rm vis}$/$E_{\rm cm}$
for labeled final states. In each figure, entries are shown for $\psi(2S)$ 
data (points with error bars), signal Monte Carlo with arbitrary normalization
(open histogram), and scaled continuum (shaded histogram). The vertical arrows indicate 
ends of signal selection ranges.} 
  \label{fig:xtot}
\end{figure}

\begin{figure}[htbp]
  \centering
  \includegraphics[height=0.7\textheight,bb=39 148 527 644,clip=true]
   {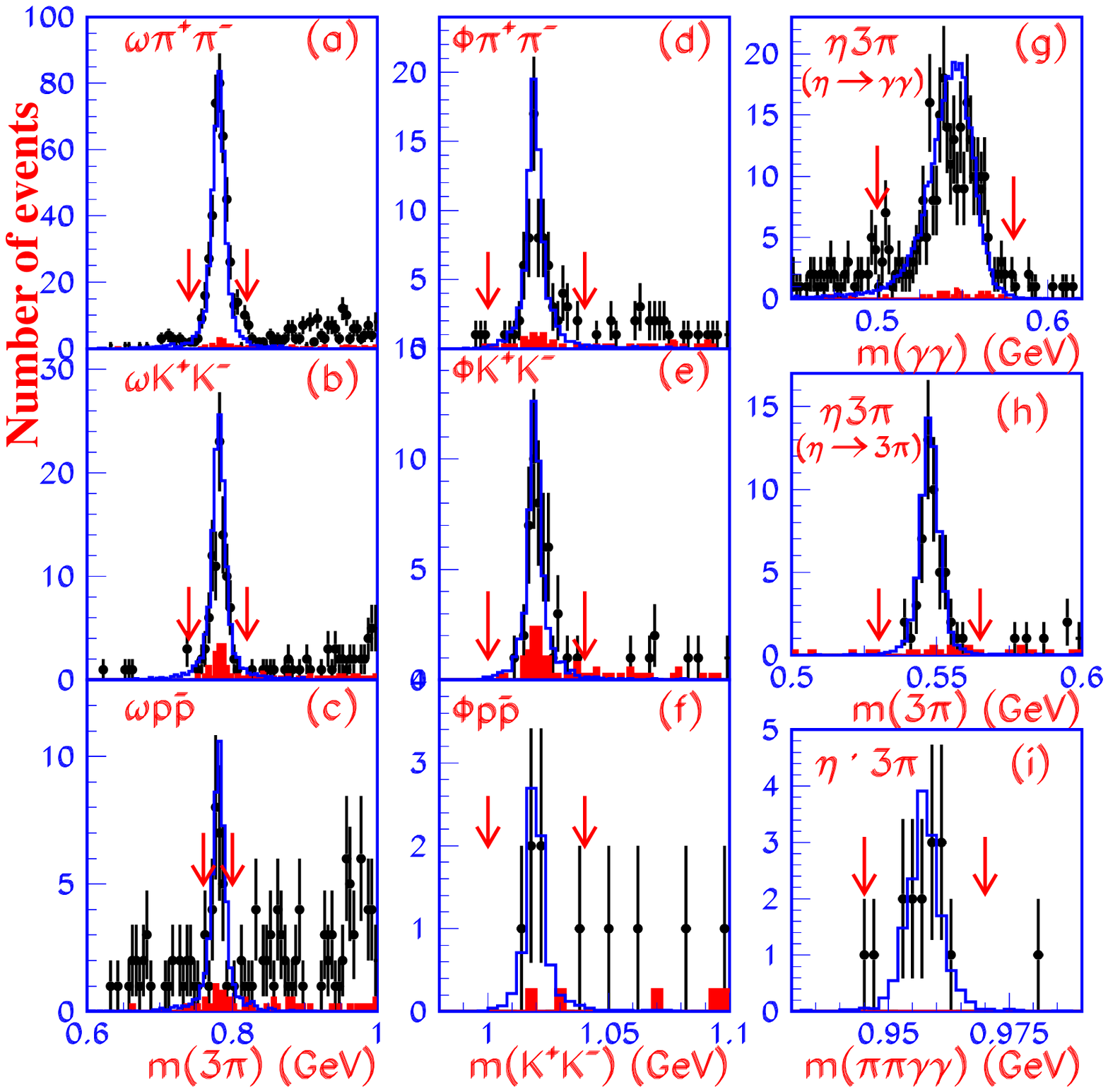}
  \caption{Distributions of the intermediate particle mass
for labeled final states. In each figure, entries are shown for $\psi(2S)$ data (points with error bars), signal Monte Carlo with arbitrary normalization
(open histogram) and scaled continuum (shaded histogram). The vertical arrows indicate 
ends of signal selection ranges.}        
  \label{fig:resm}
\end{figure}

Figures \ref{fig:xtot} and \ref{fig:resm} show the 
distributions of $E_{\rm vis}$/$E_{\rm cm}$ and resonance mass, where relevant, 
for all final states. The $\psi(2S)$ data is represented 
by the points with error bars, the open histogram is signal Monte Carlo with arbitrary normalization, 
and the shaded histogram is the scaled continuum. The arrows indicate 
the scaled energy and resonance mass selection ranges. 
The requirements of energy and momentum conservation together with particle 
identification result in small  backgrounds. The dominant background
in most modes at the $\psi(2S)$ is from the continuum.

Event totals are shown for both the $\psi(2S)$ and 
the continuum in Table \ref{tab:br}, where $S_{\psi(2S)}$ $(S_{co})$ 
is the number of events in the signal region and  $B_{\psi(2S)}$ $(B_{co})$ 
the number of events in the sideband region in $\psi(2S)$ (continuum) data. 
Under the assumption that interference between $\psi(2S)$ decay 
and continuum production of the same final state  
is absent, the number of events attributable to each $\psi(2S)$ decay mode,
$N_S$, is
 
\begin{equation}
N_S = S_{\psi(2S)} - B_{\psi(2S)} - f(S_{co} - B_{co}),
\end{equation}
where $f$ is mode dependent and listed in Table \ref{tab:br}.
We observe a statistically significant signal in all modes except $\phi p \bar p$.
The signal with the least statistical significance is $\omega p \bar p$ $(4.7 \sigma)$.

The efficiency, $\epsilon$, for each final state is the 
average obtained from MC simulations~\cite{GEANT} for both detector configurations;
the two values are typically within a few percent (relative) to each other.
No initial state radiation is included in the Monte Carlo, but final state radiation is accounted for.
The efficiencies in Table \ref{tab:br} include the branching ratios 
for intermediate final states.

\begin{table}[!h]
\caption
{
For each final state $h$: 
the number 
of events in the signal region, $S_{co}$, and 
background from sidebands, $B_{co}$, in continuum data; 
the scale  factor, $f$; 
the number of events in the signal region, $S_{\psi(2S)}$, 
and background from sidebands, $B_{\psi(2S)}$, in $\psi(2S)$ data; 
the number of events attributable to $\psi(2S)$ 
decay, $N_S$, computed according to Equation (2);
the average efficiency, $\epsilon$; 
the ratio 
${\cal B} (\psi(2S)\to h)$/${\cal B} (\psi(2S)\to\pi^+\pi^-J/\psi,
J/\psi \to \mu^+ \mu^-)$ and the absolute branching fraction with 
statistical (68\% C.L.) and systematic errors.
For $\eta 3 \pi$, the two decays modes  
$\eta \to \gamma\gamma $ 
and  $\eta \to 3 \pi$  
are  
combined on line $\eta 3\pi$.
The branching ratios reported here explicitly exclude contributions from
$\psi(2S) \to J/ \psi X$, where $X = \pi^+ \pi^-$, $\pi^0 \pi^0$, or $\eta$. 
}
\begin{center}
\begin{tabular}{|c|c|c|c|c|c|c|c|c|c|}  \hline
mode     
          &  $S_{co}$  & $B_{co}$   
                                         & $f$      
                                                 & $S_{\psi(2S)}$ & $B_{\psi(2S)}$ 
                                                              & $N_S$ 
                                                                       &$\varepsilon$
& \rule[-3mm]{0cm}{8mm}$\frac{{\cal B}(\psi(2S) \to h)}{{\cal B}^{\psi(2S)}_{\pi^+\pi^- J/\psi}{\cal B}^{J/\psi}_{\mu^+\mu^-}}$ 
                           &          ${\cal B}(\psi(2S) \to h)$  \\ 
$h$   &  & & & & & &  &  ($10^{-4}$)     &($10^{-4}$) \\ \hline
$2(\pi^+ \pi^-)$              &1437 & 39 & 0.2680 &  700 & 19 &  306.3 & 0.4317 & $ 105.1\pm 11.5\pm 13.6$ & $ 2.0\pm0.2\pm0.3$ \\
$2(\pi^+ \pi^-)\pi^0$         & 343 & 28 & 0.2549 & 1742 & 39 & 1622.7 & 0.1927 & $1247.1\pm 46.2\pm171.3$ & $23.7\pm0.6\pm3.3$ \\
$\omega \pi^+ \pi^-$          &  45 &  7 & 0.2412 &  426 & 38 &  378.8 & 0.1338 & $ 419.5\pm 33.7\pm 58.5$ & $ 8.0\pm0.5\pm1.1$ \\
$\eta^{\to\gamma\gamma} 3\pi$ 
                               &  28 &  2 & 0.2507 &  271 & 49 &  215.5 & 0.0595 & $ 536.8\pm 62.8\pm 92.6$ & $10.2\pm0.8\pm1.8$ \\
$\eta^{\to\pi^+\pi^-\pi^0} 3\pi$ 
                               &  19 & 10 & 0.2676 &   50 &  1 &   46.6 & 0.0186 & $ 370.2\pm 66.8\pm 59.1$ & $ 7.0\pm1.1\pm1.3$ \\
$\eta 3\pi$                   &     &    &        &      &    &        &        &                          & $ 8.5\pm0.7\pm1.6$ \\
$\eta^\prime 3\pi$            &   0 &  0 & 0.2349 &   14 &  2 &   12.0 & 0.0079 & $ 225.1\pm113.9\pm 69.9$ & $ 4.3\pm1.5\pm1.2$ \\
$K^+K^-\pi^+\pi^-$            & 825 &103 & 0.2694 & 1018 & 49 &  774.5 & 0.3372 & $ 340.2\pm 20.6\pm 44.0$ & $ 6.5\pm0.3\pm0.8$ \\
$\phi \pi^+ \pi^-$            &  26 & 12 & 0.2738 &   72 & 18 &   50.2 & 0.1545 & $  48.1\pm 15.7\pm  9.9$ & $ 0.9\pm0.2\pm0.2$ \\
$\omega K^+ K^-$              &  57 & 11 & 0.2091 &   92 &  8 &   74.4 & 0.1110 & $  99.3\pm 17.8\pm 14.8$ & $ 1.9\pm0.3\pm0.3$ \\
$2(K^+K^-)$                   & 100 & 10 & 0.2727 &   81 &  2 &   54.5 & 0.2605 & $  31.0\pm  8.4\pm  4.6$ & $ 0.6\pm0.1\pm0.1$ \\
$\phi K^+ K^-$                &  45 & 16 & 0.2729 &   46 &  4 &   34.1 & 0.1331 & $  37.9\pm 12.2\pm  6.8$ & $ 0.7\pm0.2\pm0.1$ \\
$p \bar{p} \pi^+ \pi^-$       & 328 & 31 & 0.2552 &  963 & 31 &  856.2 & 0.4481 & $ 283.0\pm 14.5\pm 36.5$ & $ 5.4\pm0.2\pm0.7$ \\
$\omega p \bar{p}$            &  18 &  5 & 0.2314 &   29 &  8 &   18.0 & 0.1060 & $  25.1\pm 11.4\pm  6.8$ & $ 0.5\pm0.2\pm0.1$ \\
$p \bar{p} K^+ K^-$           &  17 &  0 & 0.2563 &   29 &  1 &   23.6 & 0.3445 & $  10.2\pm  3.2\pm  1.7$ & $ 0.2\pm0.1\pm0.1$ \\

%
$\phi p \bar{p}$ & 2 & 1 & 0.2472 & 6 & 2 & 3.8 & 0.1242 & & $<0.18$ (90\%CL) \\

$\Lambda \bar\Lambda$         &   1 &  0 & 0.2431 &  112 &  0 &  111.8 & 0.1035 & $ 160.0\pm 19.9\pm 22.8$ & $ 3.0\pm0.3\pm0.4$ \\
$\Lambda\bar\Lambda\pi^+\pi^-$&   9 &  2 & 0.1775 &   46 &  3 &   41.8 & 0.0437 & $ 141.4\pm 27.1\pm 26.6$ & $ 2.7\pm0.5\pm0.6$ \\
$\Lambda \bar{p} K^+$         &  38 &  5 & 0.2258 &   51 &  8 &   35.5 & 0.1429 & $  36.8\pm  9.9\pm  6.3$ & $ 0.7\pm0.2\pm0.1$ \\
$\Lambda\bar{p}K^+\pi^+\pi^-$ &  17 &  1 & 0.1370 &   26 &  2 &   21.8 & 0.0514 & $  62.8\pm 23.6\pm 15.0$ & $ 1.2\pm0.3\pm0.3$ \\ \hline
\end{tabular}
\label{tab:br}
\end{center}
\end{table}

We correct $N_S$ for the efficiency and normalize to the  number of
$\psi(2S)\to\pi^+\pi^-J/\psi$, $J/\psi\to\mu^+\mu^-$ decays
in the data, which has been determined previously to be $(6.75 \pm 0.12)\times 10^4$,
where the error accounts for data and MC statistics (0.7\%)
and for uncertainties in trigger efficiency (0.5\%) and angular distributions (1.5\%).
Complete details may be found in~\cite{BHHMK}.
The resulting relative  branching ratios are listed in Table \ref{tab:br}.
The absolute branching ratios are
determined using 
${\cal B} (\psi(2S)\to\pi^+\pi^-J/\psi) =  0.323 \pm 0.013$ 
~\cite{PDG} and 
${\cal B} (J/\psi\to\mu^+\mu^-)=  (5.88 \pm 0.10)\% $~\cite{PDG}. 
$Q_h$ values are determined using the absolute $\psi(2S)$ branching ratios
determined in this analysis and $J/\psi$ branching ratios from~\cite{PDG}.

The systematic errors on the ratio of branching fractions
share common contributions from the number of
$\psi(2S)\to\pi^+\pi^-J/\psi$, $J/\psi\to\mu^+\mu^-$ decays (1.7\%), 
uncertainty in $f$
(1.5\%), trigger efficiency (1\%), electron veto (1\%), and Monte Carlo 
statistics (2\%). Other sources vary by channel. We include the
following contributions for 
detector performance modeling quality:
charged particle tracking (3\% per track),
$\pi^0/\eta$ finding (4.4\%), $\Lambda$ finding (3\%),
$\pi / K/ p$ identification $(1\%/ {\rm ~identified~} \pi / K /p)$, and
scaled energy and mass resolutions (2\%).
The systematic error associated with the uncertainty in the level of background 
is obtained by recomputing the branching ratio when the background 
at the $\psi(2S)$ and the continuum are coherently increased by 
$1\sigma$. Since the background in many modes is small, the Poisson 
probability for the observed number of background events to fluctuate  
up to the 68\% C.L. value is calculated and interpreted 
as the uncertainty in the level of background.
The efficiencies were determined using events generated according to phase space.
Many of the modes studied may have resonant submodes. 
A preliminary analysis finds the following prominent examples: 
$2(\pi^+ \pi^-)$ where $\rho \pi \pi$ is dominant, 
$\omega \pi \pi$ where $b_1 \pi$ is dominant and $\omega f_2(1270)$ is significant, 
$K^+K^-\pi^+\pi^-$ where $K^*(892) K \pi$ and $\rho K^+ K^-$ are large
and   $\phi \pi^+ \pi^-$ is small, and 
$\eta 3\pi$ where  $\eta \pi^0 \rho^0$, $\eta \rho^+ \pi^-$, and $\eta \rho^- \pi^+$
dominate. 
Allowing for the presence of resonant submodes changes the efficiency by less
than 
5\% relative to the non-resonant efficiency for each mode we have studied.    
To account for this and the possibility of resonance polarization,
we assign a 10\% modeling systematic error to all modes. 
Systematic uncertainties
are significant for all modes and the dominant error for many.

\begin{figure}[htbp]
  \centering
  \includegraphics[height=0.7\textheight,bb=75 156 521 644,clip=true]
   {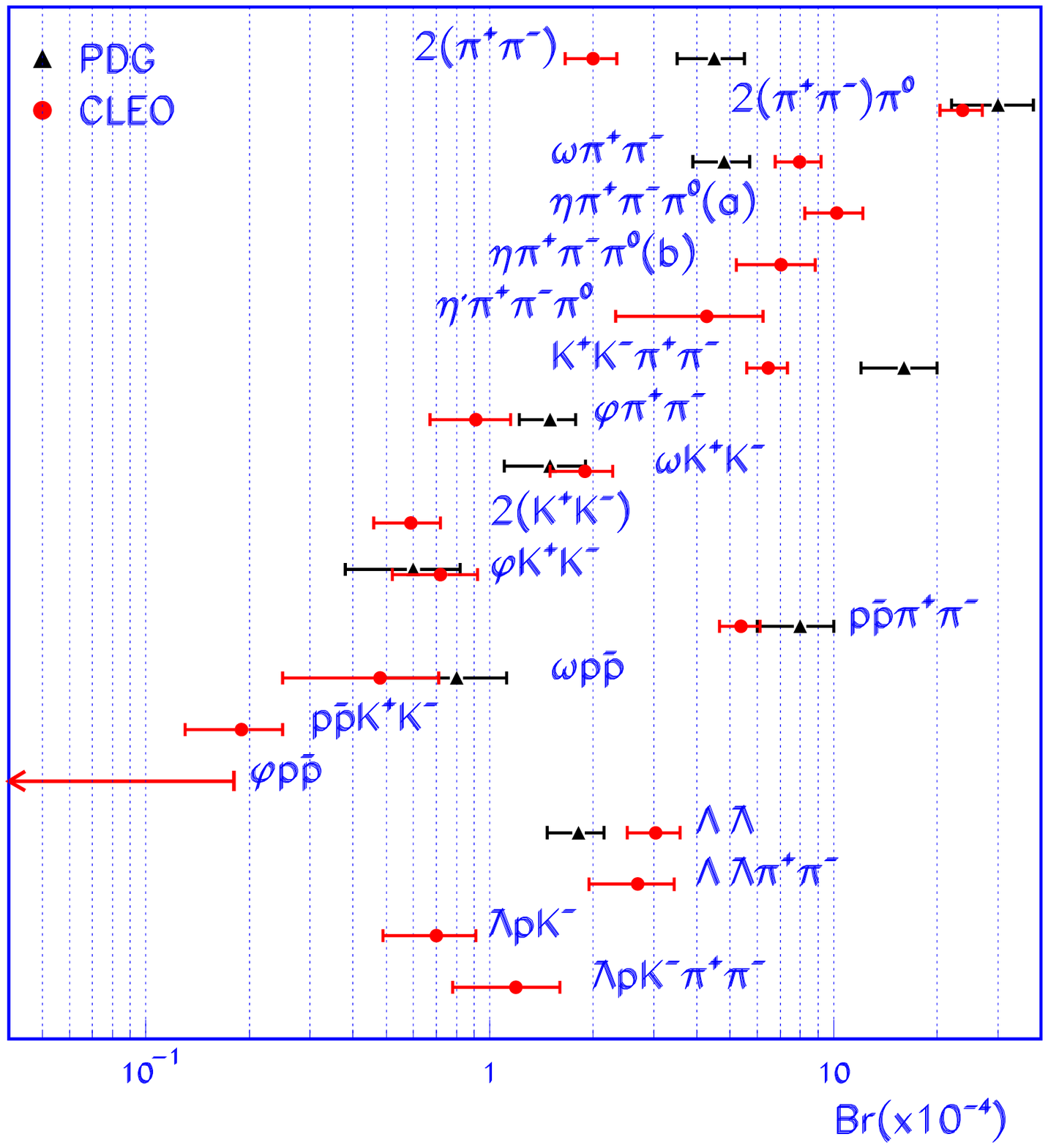}
  \caption
{
A comparison of the measurements of the $\psi(2S)$ branching ratios for the labeled 
final states from this work (circles) and, when available, 
from the PDG~\cite{PDG} (triangles). The error bars are the quadrature sum of
the statistical and systematic errors. 
}
  \label{fig:br}
\end{figure}

\begin{figure}[htbp]
  \centering
  \includegraphics[height=0.7\textheight,bb=72 153 528 644,clip=true]
   {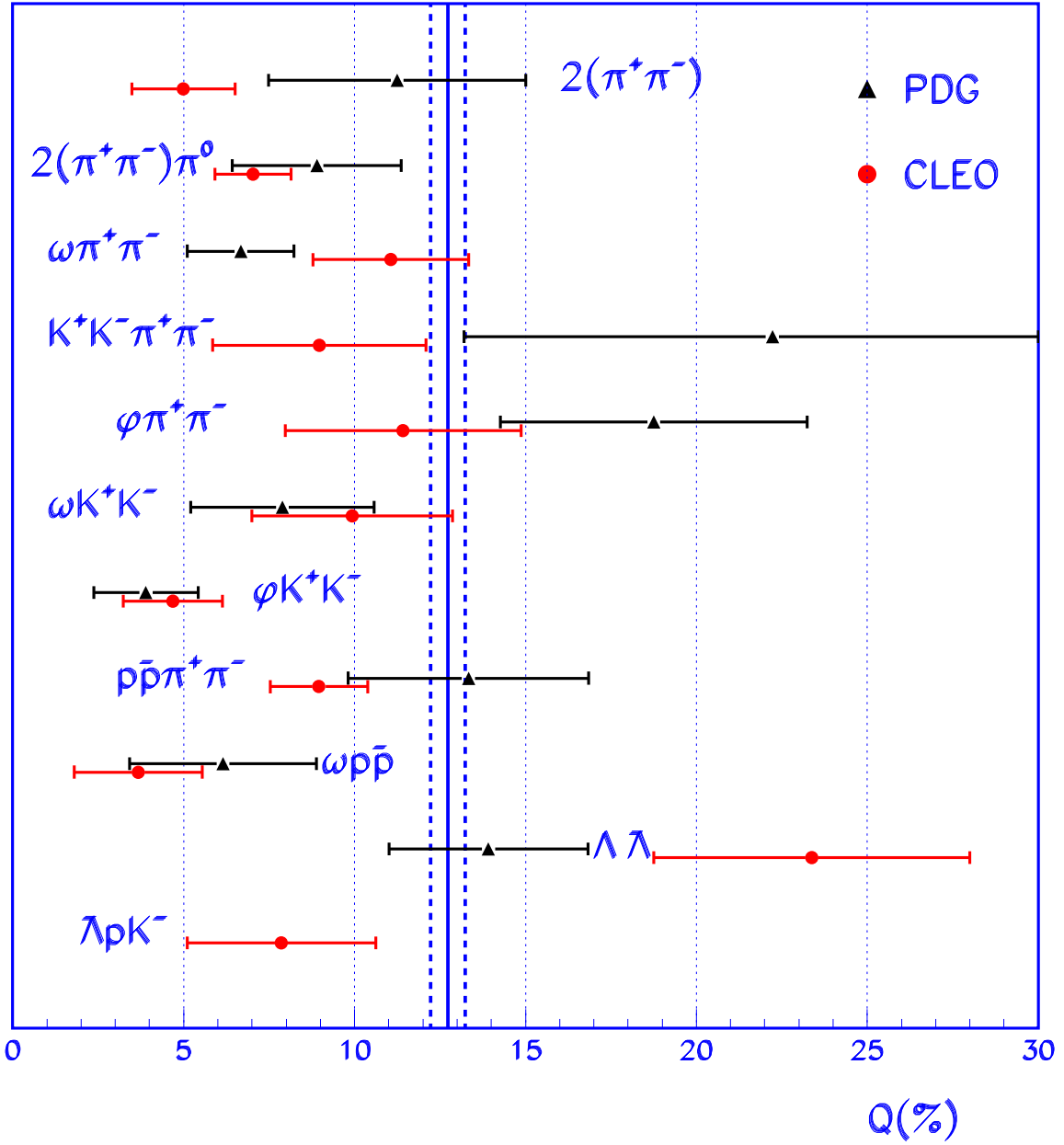}
  \caption
{
A comparison of $Q_h$ values for the labeled 
final states determined from the $\psi(2S)$ branching ratios
measured in this work (circles) and $J/\psi$ branching ratios from the PDG~\cite{PDG}
to $Q_h$ values
determined from the PDG~\cite{PDG} (triangles). The error bars are the quadrature sum of
the statistical and systematic errors. The current value 
of $Q_l=(12.7\pm 0.5) \%$~\cite{PDG}
is displayed as the vertical solid line. The vertical dashed lines
are the $\pm 1 \sigma$ uncertainty in $Q_l$.  
}
  \label{fig:q}
\end{figure}

Figure \ref{fig:br} shows the branching ratios measured 
in this analysis and a comparison to previous measurements~\cite{PDG}.
In Table \ref{tab:q} the values of $Q_h$ for those final states with measured 
branching ratios at the $J/\psi$~\cite{PDG} are given.
Figure \ref{fig:q} shows the $Q_h$ values from Table \ref{tab:q} 
and a comparison to  previous measurements.
This is the first measurement of $Q_{\Lambda \bar p K}$.
The values of $Q_h$ appear to be independent of the final state
with no significant differences between mesonic and baryonic modes.
The data indicates that $Q_h$ values are in general lower than the current leptonic 
ratio.

\begin{table}[!h]
\caption
{
For each final state $h$: the branching ratio from Table III,
the corresponding branching ratio at the $J/\psi$ from~\cite{PDG}, and $Q_h$ from Equation (1).
The $\psi(2S)$ branching ratios and $Q_h$ explicitly exclude contributions from
$\psi(2S) \to J/ \psi X$ where $X = \pi^+ \pi^-$, $\pi^0 \pi^0$, or $\eta$.
}
\begin{center}
\begin{tabular}{|c|c|c||c|c|}  \hline
   mode             &\multicolumn{2}{c||}{${\cal B}(\psi(2S)\to h) \ (10^{-4})$} & $Br(J/\psi)$ $(10^{-4})$ & $Q_h$ (\%)\\ 
   $h$ &this work & PDG  & PDG & this work \\ \hline
$2(\pi^+ \pi^-)$              &  2.0 $\pm$ 0.3 & $4.5 \pm 1.0$ &  40    $\pm$ 10    &   5.0 $\pm$1.5\\
$2(\pi^+ \pi^-)\pi^0$         & 23.7 $\pm$ 3.3 & $ 30 \pm  8 $ & 337    $\pm$ 26    &   7.0 $\pm$1.1\\
$\omega \pi^+ \pi^-$          &  8.0 $\pm$ 1.2 & $4.8 \pm 0.9$ &  72    $\pm$ 10    &  11.1 $\pm$2.3\\
$\eta 3\pi$                   &  8.5 $\pm$ 1.0 & -             &          -         &          -      \\
$\eta^\prime 3\pi$            &  4.3 $\pm$ 2.0 & -             &          -         &          -      \\
$K^+K^-\pi^+\pi^-$            &  6.5 $\pm$ 0.9 & $ 16 \pm  4 $ &  72    $\pm$ 23    &   9.0 $\pm$3.1\\
$\phi \pi^+ \pi^-$            &  0.9 $\pm$ 0.2 & $1.50\pm0.28$ &   8.0  $\pm$  1.2  &  11.4 $\pm$3.5\\
$\omega K^+ K^-$              &  1.9 $\pm$ 0.4 & $1.5 \pm 0.4$ &  19    $\pm$  4    &   9.9 $\pm$2.9\\
$2(K^+K^-)$                   &  0.6 $\pm$ 0.1 & -             &          -         &          -      \\
$\phi K^+ K^-$                &  0.7 $\pm$ 0.2 & $0.60\pm0.22$ &  15.4  $\pm$  2.1  &   4.7 $\pm$1.6\\
$p \bar{p} \pi^+ \pi^-$       &  5.4 $\pm$ 0.7 & $8.0 \pm 2.0$ &  60    $\pm$  5    &   9.0 $\pm$1.4\\
$\omega p \bar{p}$            &  0.5 $\pm$ 0.2 & $0.80\pm0.32$ &  13    $\pm$  2.5  &   3.7 $\pm$1.9\\
$p \bar{p} K^+ K^-$           &  0.2 $\pm$ 0.1 & -             &          -         &          -      \\
$\phi p \bar{p}$              & $<0.18$(90\%CL)& $<0.26$       &   0.45 $\pm$  0.15 &          -      \\
$\Lambda \bar\Lambda$         &  3.0 $\pm$ 0.5 & $1.81\pm0.34$ &  13    $\pm$  1.2  &  23.4 $\pm$4.6\\
$\Lambda\bar\Lambda\pi^+\pi^-$&  2.7 $\pm$ 0.8 & -             &          -         &          -      \\
$\Lambda \bar{p} K^+$         &  0.7 $\pm$ 0.2 & -             &   8.9  $\pm$  1.6  &   7.9 $\pm$2.8\\
$\Lambda\bar{p}K^+\pi^+\pi^-$ &  1.2 $\pm$ 0.4 & -             &          -         &          -      \\ \hline
\end{tabular}
\label{tab:q}
\end{center}
\end{table}

In summary, we have presented preliminary branching fractions 
for seven new decay modes of the $\psi(2S)$, namely
$\eta 3\pi$, $\eta^\prime 3\pi$, $2(K^+ K^-)$, $p \bar p K^+ K^-$,
$\Lambda \bar \Lambda \pi^+ \pi^-$, $\Lambda \bar p K^+$,
$\Lambda \bar p  K^+ \pi^+ \pi^-$,
and
more precise measurements of nine previously measured modes, which are
$2(\pi^+ \pi^-)$, $2(\pi^+ \pi^-) \pi^0$, $\omega \pi^+ \pi^-$,
$K^+ K^- \pi^+ \pi^-$, $\phi \pi^+ \pi^-$, $\omega K^+ K^-$, $\phi K^+ K^-$,
$p \bar p \pi^+ \pi^-$, and $\Lambda \bar \Lambda$. 
We also measure $ \omega p \bar p $
and obtain an improved upper limit for $\phi p \bar p$.
Results are compared, where possible, with the corresponding $J/\psi$ branching ratios to test the
12\% rule.  This analysis is part of a 
comprehensive study of  $\psi(2S)$  multi-body hadronic decays and 
their resonant substructure and is closely related  
to searches for  non-$D\bar D$ decay modes of the $\psi(3770)$. Further results 
will be presented in the near future.   

We gratefully acknowledge the effort of the CESR staff 
in providing us with
excellent luminosity and running conditions.
This work was supported by 
the National Science Foundation,
the U.S. Department of Energy,
the Research Corporation,
and the Texas Advanced Research Program.


\begin{thebibliography}{99}
\bibitem{note} We do not veto $J/\psi$ from $\psi(2S) \rightarrow J/\psi \pi^0$ 
since it is a very small contribution as
${\cal B}(\psi(2S) \rightarrow J/\psi \pi^0)= (9.6 \pm 2.1) \times 10^{-4}$.
\end{thebibliography}
\end{document}